\documentclass[10pt, aps, prl, twocolumn, superscriptaddress, floatfix, showpacs]{revtex4-2}
\usepackage{graphicx}
\graphicspath{{Figures/}}
\usepackage{amsfonts}
\usepackage{amssymb}
\usepackage{amsmath}
\usepackage{color}
\usepackage{wasysym}
\usepackage{hyperref}
\usepackage{bbold}
\usepackage{algpseudocode}
\usepackage{placeins}

\definecolor{lightblue}{RGB}{136,163,209}

\begin{abstract}
  The optimization of the power consumption of antenna networks is a problem with a potential impact in the field of telecommunications. In this work, we investigate the application of the quantum approximate optimization algorithm (QAOA) and the quantum adiabatic algorithm (QAA), to the solution of a prototypical model in this field.
  We use statevector emulation in a high-performance computing environment to compare the performance of these two algorithms in terms of solution quality, using selected evaluation metrics.
  We estimate the circuit depth scaling with the problem size while maintaining a certain level of solution quality, and we extend our analysis up to 31 qubits, which is rarely addressed in the literature.
  Our calculations show that as the problem size increases, the probability of measuring the exact solution decreases exponentially for both algorithms. This issue is particularly severe when we include constraints in the problem, resulting in full connectivity between the sites. Nonetheless, we observe that the cumulative probability of measuring solutions close to the optimal one remains high also for the largest instances considered in this work.
  Our findings keep the way open to the application of these algorithms, or variants thereof, to generate suboptimal solutions at scales relevant to industrial use-cases. 
\end{abstract}

\begin{document}

\title{Evaluating the Practicality of Quantum Optimization Algorithms for Prototypical  Industrial Applications}

\author{M. Vandelli}
\affiliation{High-Performance Computing Research Laboratory, Leonardo S.p.A., Via R. Pieragostini 80, Genova, Italy}
\author{A. Lignarolo}
\affiliation{High-Performance Computing Research Laboratory, Leonardo S.p.A., Via R. Pieragostini 80, Genova, Italy}
\author{C. Cavazzoni}
\affiliation{High-Performance Computing Research Laboratory, Leonardo S.p.A., Via R. Pieragostini 80, Genova, Italy}
\author{D. Dragoni}
\affiliation{High-Performance Computing Research Laboratory, Leonardo S.p.A., Via R. Pieragostini 80, Genova, Italy}
\maketitle

\section{Introduction}

Combinatorial optimization problems are ubiquitous in industry, science, and technology. A particularly interesting class of problems is the binary quadratic problem \cite{Kochenberger2014}, which can be equivalently written in the QUBO (quadratic unconstrained optimization problem) or Ising forms \cite{Lucas_2014, lodewijks2020mapping}. Many representatives of this class of problems are NP-hard to solve on classical digital computers \cite{FBarahona_1982}. Consequently, quantum workflows could provide advantages with respect to traditional classical approaches \cite{farhi2019quantum} in terms of speed and accuracy, once a sufficient number of qubits becomes available \cite{Guerreschi2019}. An efficient solution to these optimization problems can potentially have a beneficial impact on many companies and is one of the reasons behind the emergence of a new market centered around quantum computing capabilities \cite{MacQuarrie2020}.

Most of the previous analyses involving the application of quantum algorithms to QUBO problems focused on problems of academic interest, such as the MaxCut problem \cite{PhysRevA.103.042612} and the Sherrington-Kirkpatrick model \cite{Farhi_2022}. Although useful in demonstrating the advantages and drawbacks of quantum computing methods, these analyses do not consider some issues that characterize real-world optimization problems. The latter are generally unstructured, characterized by continuous weights and often exhibit a high connectivity of the associated graph, especially in the presence of constraints. Even though these problems are arguably harder to solve even for a quantum computer \cite{PhysRevX.10.021067, koßmann2023deepcircuit}, stringent bounds on the performances of quantum algorithms for general problems have not been derived yet. Another important requirement in real-world problems is that a good value of the cost function must be accompanied by the configuration corresponding to that cost to be of practical utility. 

So far, problems of industrial interest were mainly studied with quantum annealing techniques \cite{PhysRevE.58.5355,Johnson2011,Boixo2014}, which is specifically tailored to tackle this kind of problems \cite{smelyanskiy2012nearterm,Rieffel2015,venturelli2016quantum, 10.1145/3293320.3293333, Stollenwerk_2020, Yarkoni_2022,Domino_2022}, including optimization of antenna networks \cite{9863923,9884342,Colucci2023}. Only recently some real-world problems were investigated using universal gate-based quantum computing techniques. The list includes power grid optimization \cite{Jing2023}, the flight-gate assignment problem \cite{Stollenwerk2020TowardQG}, the vehicle routing problem \cite{leonidas2023qubit}, and the satellite mission planning problem \cite{quetschlich2023hybrid,Rainjonneau_2023} just to name a few.

\section{Simplified Ising model for the antenna placement problem}

\begin{figure*}[t!]
    \centering
\includegraphics[width=\textwidth]{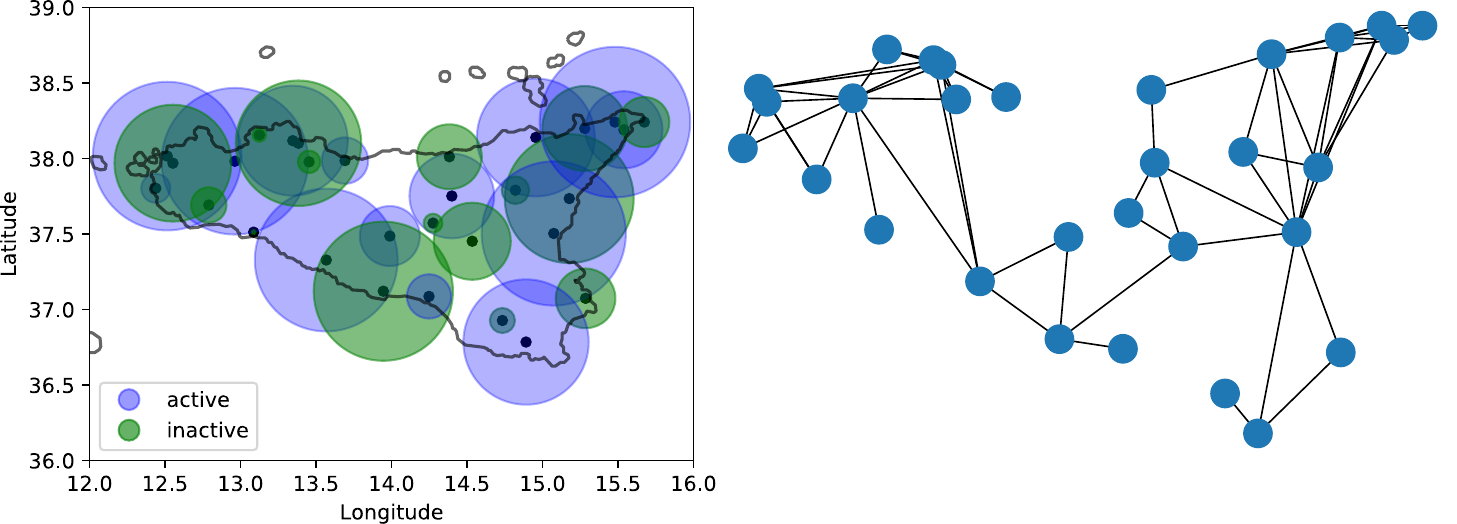} 
    \caption{Left panel: optimal solution of a selected antenna problem with $N=30$ sites and the soft penalty factor $\lambda=0$. Black dots, arbitrarily distributed on Sicily island to connect with a concrete geographical area, indicate the location of each antenna. Green circles correspond to active antennas, while blue circles indicate inactive ones. Right panel: the corresponding graph describing the connectivity of the considered problem. Each node indicates an antenna $i$ with weight $-\xi A_i$ and each edge represents a non-zero matrix element of the connectivity matrix $J_{ij}$.}
    \label{fig:antenna_location}
\end{figure*}

The aim of this work is the study of a simplified model for the optimization of a network of antennas. Our model is relevant in an emergency scenario characterized by a failure of the power grid and by difficulties in providing supplies to remote areas. In such situations, establishing reliable communication is essential to ensure that rescuers can communicate with each other, quickly coordinate with the control room, and respond to the crisis effectively \cite{electronics11071155}. A lack of high-performance reliable and secure communications can potentially have negative consequences. For instance, the antennas needed for emergency communications can be static transmitters powered by electric generators, e.g. diesel generators, or they can be deployed on unmanned aerial vehicles (UAVs) used as communication bridges \cite{7486987, SUN2023}.

It is therefore critical to ensure that the signal generated by the antennas provides the maximum possible coverage of the territory while minimizing the areas where the signal overlaps with neighboring antennas. Overlaps need to be minimized in order to achieve homogeneous antenna distributions that reduce costs of operations and to avoid waste of local resources required by each antenna, such as fuel or UAV batteries. In the case of UAV communication bridges, the coverage redundancy can result in an excess of transmitted data, channel interference and the need for a larger constellation of UAVs, as well as a high energy consumption that leads to a shortening of the lifespan of the multi-UAV network. Additionally, only a limited amount of antennas can be readily available to be dispatched on the territory, especially if a disruption of the logistic infrastructure occurred. 

To deal with a small number of parameters, we adopt the following model. We consider a single instance of this problem by placing $N_{\rm t}$ operating antennas at $N$ potential locations corresponding to the geographical coordinates of cities in Sicily. Each potential site $i$ is characterized by a binary label describing its status, active ($z_i=+1$) or inactive ($z_i=-1$).
The Ising cost function of the problem can be written in terms of strings of binary variables $z \in \mathcal{S}_N = \{-1, +1\}^{\otimes N}$, as a sum of an overlap term $H_O$, a coverage term $H_C$, and a soft-penalty term $H_P$ as
\begin{align}
    H(z; \; \xi, \lambda, N_{\rm t}) = H_O(z) +  H_C(z) +  H_P(z) 
\end{align}
with
{\allowdisplaybreaks
\begin{align}
    &H_O(z) = \sum_{i \neq j} z_i \, J_{ij} z_j\\
    &H_C(z) = - \xi \sum_i A_i z_i\\
    &  H_P(z) = \lambda\left(\sum_{i \neq j}  z_i z_j + \delta N_{\rm t}\sum_i z_i\right).
\end{align}
}
The solution to the problem is given by the string $z_{\rm gs}$, which minimizes the cost function and we define the minimum cost as $H_{\rm min} = H(z_{\rm gs})$. We call the string $z_{\rm gs}$ as the ground state in analogy with physical systems. The chosen set of possible locations of the antennas is shown in Fig.\ref{fig:antenna_location}. The black dots are the precise locations, while the circles indicate the area covered by each antenna. The colors indicate which locations are operative (blue circles) or turned off (green circles).

The parameters $J_{ij}$ and $A_i$ represent areas. Considering the circle $B_i$ centered at antenna $i$, we define the overlap coefficients between antennas located at positions $i$ and $j$ as $J_{ij} = {\rm Area}(B_i \cap B_j)$, while the coefficients of the linear term correspond to its area $A_i = {\rm Area}(B_i)$. The radius of each circle is randomly generated according to the uniform distribution up to a maximum value. The positive dimensionless numbers $\xi$ and $\lambda$ can be tuned to modify the relative weight of the two terms. The number $\lambda$ introduces a soft penalty for all the configurations with the wrong number of operating antennas and must be chosen large enough to ensure that the ground state has the correct number of antennas and $\delta N_{\rm t} = N-2N_{\rm t}$. This expression for the number constraint can be derived by considering that the network must have $N_{\rm t}$ active sites with $z=+1$ and $N-N_{\rm t}$ inactive sites with $z=-1$, hence the constraint reads $\left(\sum_i z_i\right) = N_{\rm t} - (N-N_{\rm t})$. 

When $\lambda>0$, the $H_P$ term introduces connectivity between all the sites, regardless of their distance. Although simplified, this model establishes a practical and essential connection to reality and can be readily extended or adapted to a specific real scenario as required.  
In short, the problem is described by the cost function
\begin{align}
   H(z; \xi, \lambda, N_{\rm t}) = \sum_{i\neq j}  \tilde{J}_{ij} z_i  z_j + \sum_{i} \tilde{A}_i z_i  
\end{align}
with
\begin{align}
  \tilde{J}_{ij} = J_{ij} + 2\lambda\delta N_{\rm t} \\
  \tilde{A}_i = 2 \lambda\delta N_{\rm t}-\xi A_i
\end{align}
In the rest of this work, we consider the cases $\lambda = 0$ and $\lambda > 0$ with $N_t=\lfloor N/2 \rfloor$. Additionally, we have to choose a value for $\xi$. A choice of $\xi \in (0, 1)$ gives more importance to reducing the overlap, so we fix $\xi = 1/4$ to ensure a good balance between overlap and coverage for this specific problem. The associated cost function is shortly identified by ${H_\lambda(z) = H(z; \xi=1/4, \lambda, N_{\rm t}=\lfloor N/2 \rfloor)}$. The Ising problem can be represented as a graph, in which each edge connecting node $i$ to node $j$ is associated with the matrix element $J_{ij}$, and each node $i$ has weight $A_i$, as shown in the lower panel of Fig.\ref{fig:antenna_location}. This Ising problem has a nonhomogeneous local field, and the corresponding graph is unstructured, weighted, and nonplanar. If $\lambda=0$, the graph is planar for $N < 25$. These properties make it NP-hard to solve \cite{FBarahona_1982}. This statement can also be justified by noticing that this problem is equivalent to a weighted MaxCut problem \cite{doi:10.1287/opre.36.3.493}. 

\begin{figure}[t!]
    \centering
        \includegraphics[width=0.47\textwidth]{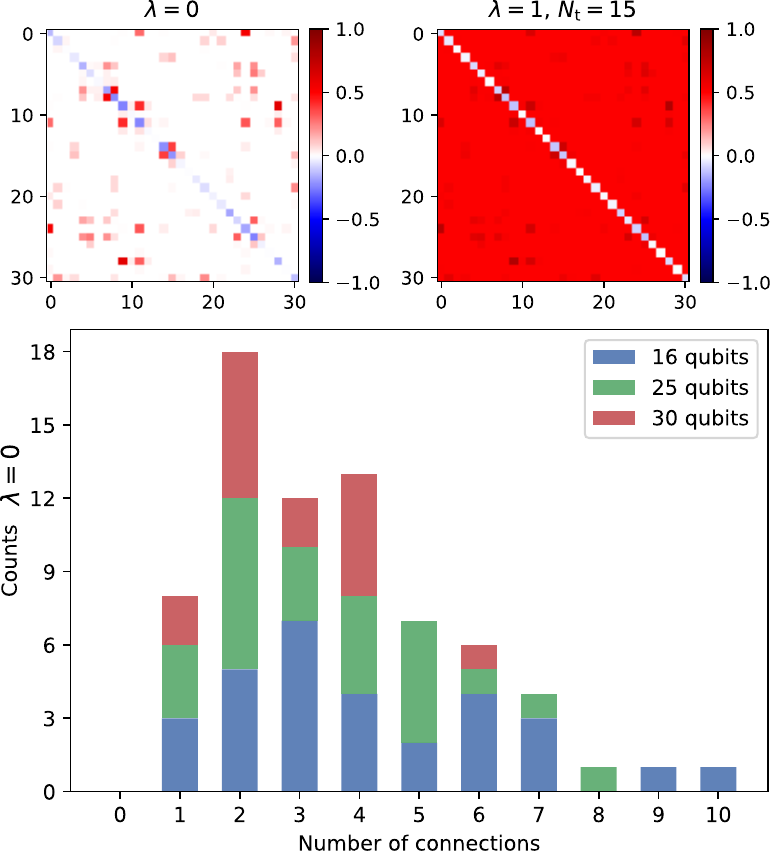}
    \caption{Upper panel: Normalized interaction matrix defined as $C_{ij} = \tilde{J}_{ij}$ for $i \neq j$ and  $C_{ii} = \tilde{A}_i$ of the chosen antenna problem with $\lambda =0$ (left) and $\lambda=1$ with $N_{\rm t} = 15$ (right). Lower panel: histogram of the connectivity that counts how many antennas have a given number of connections for different total numbers of sites. }
    \label{fig:characterization_problem}
\end{figure}

In the case $\lambda = 0$, the number of antennas can vary freely depending on which configuration minimizes the other cost terms. Instead, the $\lambda>0$ case favors those solutions with the preselected population $N_{\rm t}$ of antennas on the available sites. 
Although the two problems can be reduced to the same parametric cost function, the constraint introduces a difference in the connectivity of the problem, i.e. the number of non-zero elements of the matrix $\tilde{J}_{ij}$. In the case without number constraint, the connectivity is limited due to the local nature of the interactions, and the corresponding $J_{ij}$ is a sparse matrix. To illustrate this point, in Fig. \ref{fig:characterization_problem}, we show the interaction matrix $\tilde{J}_{ij}$ for $\lambda=0$ and $\lambda=1$. For simplicity in the visualization, we put the elements of $\tilde{A}_i$ on the diagonal (since obviously $\tilde{J}_{ii} = 0$). Since the graph is non-regular, the connectivity has a distribution as shown in the lower panel of Figure \ref{fig:characterization_problem}. The average connectivity for the $16 \leq N \leq 30$ problem is between 4 and 5. 
On the other hand, in the case $\lambda > 0$, the constraint introduces full connectivity between all the nodes of the graph. In the present work, we are explicitly interested in how the performances of quantum algorithms vary in cases with sparse and full connectivity, so we treat the constraint as a soft constraint. However, we point out that some kinds of hard constraints can be directly incorporated in QAOA and QAA via fermionic encoding \cite{PhysRevResearch.5.023071}.

\section{Methods}

\subsection{Quantum approximate optimization algorithm}

The Ising formulation presented in the previous section can be directly encoded on a quantum computer by simply promoting the classical $z_i$ variables describing the sites to quantum operators acting on a $2^N$-dimensional Hilbert space spanned by the basis of strings $\{|z\rangle\}_{z\in \mathcal{S}_N}$.
The corresponding Hamiltonian operator $\hat{H}$ acting in this Hilbert space is diagonal in the computational basis and satisfies
\begin{align}
    \hat{H}_\lambda |z\rangle = H_\lambda(z) |z\rangle,
\end{align}
hence the ground state $|z_{\rm gs}\rangle$ of $\hat{H}$ coincides with the solution $z^*$ of the classical Ising problem.
Hence, each antenna is mapped on a single qubit.

For this task, we use the quantum approximate optimization algorithm (QAOA)\cite{farhi2014quantum}, a variational quantum eigensolver (VQE) \cite{Yuan2019theoryofvariational, Cerezo2021} extensively studied in the context of combinatorial optimization problem. The QAOA uses the following parametric ansatz for the variational quantum state
\begin{align}
    |\boldsymbol{\beta}, \boldsymbol{\gamma}\rangle &= U_p(\boldsymbol{\beta}, \boldsymbol{\gamma}) |+\rangle^{\otimes N} \notag \\&=  \left(\prod_{k=1}^p e^{-i\beta_k \sum_j \hat{X}_j} e^{-i\gamma_k \hat{H}}\right) |+\rangle^{\otimes N}
\end{align}
where $\boldsymbol{\beta} = \{\beta_k\}_{k=1,...,p}$ and $\boldsymbol{\gamma} = \{\gamma_k\}_{k=1,...,p}$ are the variational parameters and $\hat{X}_j$ is the first Pauli matrix acting on qubit $j$. The initial state is a direct product of the states $|+\rangle = \left(|0\rangle + |1\rangle\right)/\sqrt{2}$ for each qubit and $p$ is the number of layers that are applied consecutively, also known as depth of the circuit. 
The variational principle states that the ground state energy $H_{\rm min}$ satisfies
\begin{align}
    H_{\rm min} \leq \langle\boldsymbol{\beta}, \boldsymbol{\gamma}| \hat{H}_\lambda|\boldsymbol{\beta}, \boldsymbol{\gamma}\rangle,
\end{align}
so, if the ansatz is sufficiently expressive, a minimization of the expectation value $\langle\boldsymbol{\beta}, \boldsymbol{\gamma}| \hat{H}_\lambda|\boldsymbol{\beta}, \boldsymbol{\gamma}\rangle$ should lead to a good approximation to $H_{\rm min}$ characterized by a state whose probability distribution has a large weight at the optimal string \cite{Larkin_2022}.
To this aim, the set of parametric angles $\boldsymbol{\beta}$ and $\boldsymbol{\gamma}$ is optimized using classical approaches on digital hardware. The QAOA ansatz is able to span the entire Hilbert space of the $N$-qubit system, provided a sufficiently deep quantum circuit is considered \cite{PhysRevA.105.042415, Morales_2020, PhysRevLett.124.090504, koßmann2023deepcircuit}. The minimal depth required to effectively represent the state is related to the degree of correlation of the considered problem instance \cite{farhi2020quantum}. At finite $p$, accuracy bounds of QAOA were established for specific problems \cite{farhi2014quantum, PhysRevA.103.042612}.

Many variants of QAOA have been proposed to reduce circuit depth and increase the expressivity of the \emph{ansatz}  \cite{PhysRevA.105.042415, binkowski2023elementary, PhysRevResearch.4.013141, leontica2023exploring, blekos2023review}. However, in most of these variants, the number of continuous parameters to be optimized at a given depth $p$ increases with the number of qubits $N$, making them hardly applicable to real-world optimization problems with $N$ greater than $10^3$. A notable exception is the possibility of warm-starting QAOA from the solution of a relaxation method \cite{Egger2021warmstartingquantum, tate2023warmstarted}.

\subsection{Quantum adiabatic algorithm}

In the limit of $p \rightarrow +\infty$ and small angles, the QAOA coincides with the Trotter-Suzuki discretization of a quantum adiabatic evolution, called quantum adiabatic algorithm (QAA)\cite{crosson2014different}, for which precise convergence bounds based on the adiabatic theorem are known \cite{farhi2000quantum, PhysRevA.67.022314, doi:10.1126/science.1057726, binkowski2023elementary,  RevModPhys.90.015002}. 
We use a first-order formulation of QAA, so that the circuit structure of the applied gates coincides with QAOA, except that the angles are fixed to predetermined values.
The specific values of the $\boldsymbol{\beta}$ and $\boldsymbol{\gamma}$ angles are here determined by the linear schedule
\begin{align}
    &\beta_k = \Delta (1 - k/p) \notag \\
    &\gamma_k = \Delta k / p,
\end{align}
at depth $p$ \cite{kremenetski2021quantum}. This choice leaves two free parameters: the number of layers $p$ and the time-step $\Delta$. In practical terms, the first is mainly limited by the available computational resources, as the number of gates grows linearly with the number of layers. For ideal calculations, the main limit is related to the clock frequency, while in real quantum devices the limit is set by the coherence time and by the error spread as the number of gates increases. The parameter $\Delta$ depends on the characteristics of the problem and it is subject to two opposite constraints. On the one hand, $\Delta$ should be as small as possible, to ensure the validity of the first-order Trotterization. On the other hand, though, it should ensure that the total time $T$ of the adiabatic evolution is $T \geq C/\Delta_g^2$, where $\Delta_g$ is the gap between the ground state and the first excited state and $C$ a numerical constant. The effects of this trade-off were investigated in detail in Ref.\cite{kremenetski2021quantum} in the context of quantum chemistry. 

\subsection{Complexity of the two algorithms}

QAOA and QAA are two different quantum algorithms used for solving optimization problems. The most significant distinction between the two is that QAOA is a heuristic algorithm that can work with relatively shallow circuits, while QAA provides precise performance guarantees but requires circuit depths that may exceed the capabilities of current hardware. 

However, except for the different way the $\boldsymbol{\beta}$ and $\boldsymbol{\gamma}$ parameters are chosen, the circuits for QAOA and QAA are identical. Practically, the difference in computational complexity stems from the complexity of choosing the parameters (optimization in QAOA $vs$ fixed parameters in QAA) and in the number of layers needed to reach an acceptable solution.
 
Specifically, the computational cost of an adiabatic schedule at fixed $p$ scales as $O(p)$ while the non-convex optimization of the QAOA angles is in principle NP-hard \cite{PhysRevLett.127.120502} unless some heuristics is used to generate the parameters at depth $p$ based on the optimal parameters at lower depth \cite{PhysRevX.10.021067, PhysRevA.107.062404}. Recently, the use of algorithms for global optimization in the parameter space, such as Bayesian sampling \cite{tibaldi2023bayesian} or differential evolution \cite{carrascal2023differential} was investigated. Although these methods can outperform the use of local optimizers combined with random initialization, they typically require a very large number of function evaluations to properly explore the space of parameters.

Using heuristic strategies to choose the parameters is also motivated by the weak dependence of the QAOA optimal parameters on the dimension within a given class of problems \cite{https://doi.org/10.48550/arxiv.1812.04170}. In this paper, we use a heuristic strategy to determine the starting point for the QAOA optimization at depth $p$. 

Since the minimization in the space of parameters $(\boldsymbol{\beta}, \boldsymbol{\gamma})$ requires $N_{\rm iter}$ iterations to reach a local minimum, in QAOA the circuit of depth $p$ has to be repeated $N_{\rm iter}$ times.
For this reason, the time complexity of each algorithm is defined by the total number of applied layers $p_{\rm tot}$ to reach the solution, which is $p_{\rm tot} = p$ for the QAA and $p_{\rm tot} = N_{\rm iter} \cdot p$ for the QAOA.

\section{Results}

\subsection{Details of the calculations}

Our calculations focused on evaluating the performances of the QAOA and QAA algorithms for our simplified antenna model, keeping an eye on the issues that arise in industrial use-cases.
However, assessing the performances on the current noisy-intermediate scale quantum (NISQ) devices \cite{Preskill2018quantumcomputingin} is challenging, especially as it is difficult to disentangle the effect of the hardware noise from the intrinsic limitations of the algorithm. 
Our calculations focus on the exact emulation of the algorithm and are based on the \emph{qiskit} software package \cite{Qiskit} by IBM combined with an internally developed MPI framework for global optimization using a multi-walker strategy. The angles of the QAOA optimization are determined using the INTERP strategy \cite{PhysRevX.10.021067} with several parallel walkers generated within a hypercube centered around the main angles. In our simulations, we found that the use of several walkers leads to a more robust solution, as $p$ is increased. The local parameter minimization is performed with the COBYLA minimizer contained in the SciPy software package \cite{2020SciPy-NMeth}. COBYLA is also resilient to noise, making it a suitable choice in the presence of shot noise or environment noise on NISQ devices. For homogeneity, we fix the number of iterations of the COBYLA minimizer in QAOA to be $N_{\rm iter}=50$. Since the search is a local search around the INTERP angles, this choice for $N_{\rm iter}$ already led to a converged result at least in the range of problem sizes considered here. For the QAOA optimization with multiple-walkers, we used 32 walkers using a single GPU and 8 CPU cores each.
On the other hand, QAA calculations were done on 48 CPU cores and a single GPU for each point. All the calculations were performed on the proprietary \emph{davinci-1} cluster equipped with AMD EPYC 7402 24-Core CPUs and NVIDIA A100 GPUs. 

The metric we use to evaluate the accuracy of the quantum algorithms is the approximation ratio defined as
\begin{align}
    \alpha = \frac{\langle H_\lambda \rangle}{H_{\rm min}},
    \label{eq:approx_ratio}
\end{align}
where $\langle H_\lambda \rangle = \langle\beta^*, \gamma^*|\hat{H}_\lambda|\beta^*, \gamma^*\rangle$ is the expectation on the final quantum state $|\beta^*, \gamma^*\rangle$, obtained either by variational optimization or by adiabatic evolution \cite{PhysRevX.10.021067}. The exact solution of the problem is recovered when $\alpha=1$. 
Although the term \emph{approximation ratio} usually refers to a lower bound on the accuracy of an approximate algorithm on the instances of a given class of problems, we use this term here to indicate the accuracy on the single instance of interest.

A second crucial metric is the probability of finding the correct ground state. In the emulation setting, we can exactly compute the full probability $p(z)$ for all the strings. In this case, the expectation value of $\hat{H}_\lambda$ is $\langle H_\lambda \rangle = \sum_{z \in \mathcal{S}_N} p(z) H(z)$ with $p(z)$. However, on a real quantum computer, we have access to the state only through measurement operations. We can use the Aer emulator within \emph{qiskit} to emulate the situation in which only $N_{\rm meas}$ shots are available to sample the state.
The expectation value can be estimated using the mean estimator $\langle \widetilde{H_\lambda} \rangle = \sum_{k=0}^{N_{\rm meas}}  \frac{H(z_{k})}{N_{\rm nmeas}}$ where $z_k$ is the string randomly generated according to the probability distribution $p(z)$ at shot $k$. In the rest of this work, we use a tilde to indicate finite-shot estimators (ex. $\widetilde{\alpha}$) as opposed to quantities extracted from the full statevector. A finite-sample estimator of the probability itself is the normalized number of counts $N_{k}/N_{\rm meas}$ for a given state $k$. Specifically, we indicate $N_{\rm gs}/N_{\rm meas}$ in the case of the ground state. The presence of a finite $N_{\rm meas}$ introduces a shot noise, that makes the optimization of the angles more difficult \cite{scriva2023challenges}. A limited amount of measurements was already identified as a potential roadblock in quantum computational chemistry \cite{PhysRevResearch.4.033154}. On our cluster, we were able to use the exact emulation of the full state up to problem sizes $N < 25$, while we resorted to using the version based on sampling for larger problems. In analogy with Eq.\eqref{eq:approx_ratio}, we define also an approximation ratio $\alpha_{\rm MP}$ of a given state as
\begin{align}
    \alpha_{\rm MP} = \frac{H(z_{\rm MP})}{H_{\rm min}},
    \label{eq:approx_ratio_MP}
\end{align}
where MP indicates the most-probable state. The corresponding limited-shot quantity $\widetilde{\alpha}_{\rm MP}$ is the most-frequent state measured. In the $p\rightarrow +\infty$ with sufficiently small $\Delta$ parameter, we recover the ideal result of the adiabatic theorem, so $\alpha \rightarrow \alpha_{\rm MP}$ and $\alpha_{\rm MP} \equiv 1$ (most-probable state is the ground state).

\begin{figure}[t!]
    \centering
    \includegraphics[width=0.5\textwidth]{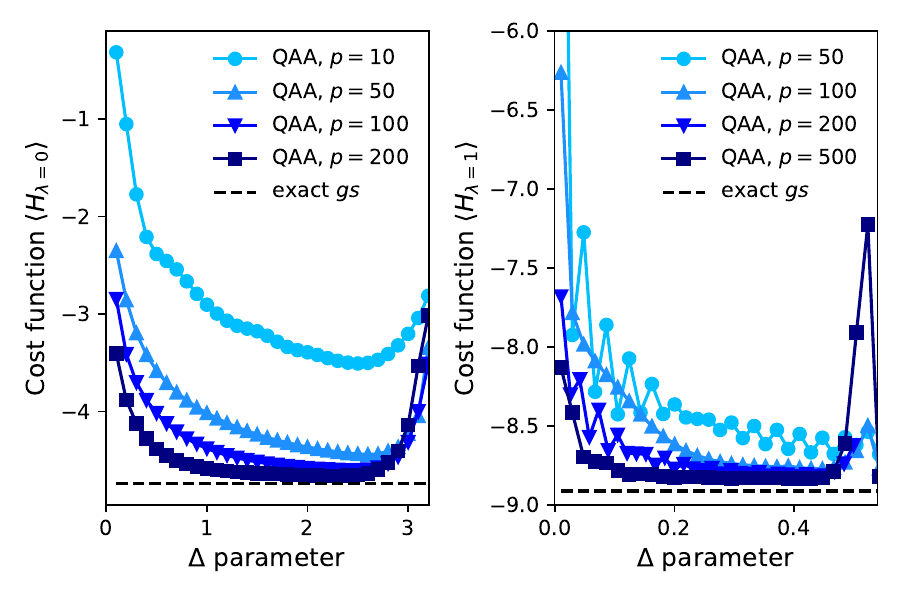}
    \caption{Cost function as a function of the  value of $\Delta$ parameter obtained from QAA using the full statevector with the linear schedule at $N=18$. Different colors indicate different circuit depths $p$. The Left panel shows the case $\lambda=0$, right panel $\lambda=1$. The dashed line indicates the exact solution to the problem instance.}
    \label{fig:qaa_vs_delta}
\end{figure}

Another important remark concerns the determination of the optimal $\Delta$ for QAA calculations. It can be obtained by varying $\Delta$ in a suitable range, as shown in Fig.\ref{fig:qaa_vs_delta}.
We observe the expected non-monotonic behavior with a minimum in the $\langle H_\lambda \rangle$ as a function of $\Delta$, at least when $p$ is suitably large. We choose $\Delta$ in the region of parameters where the cost function is minimized.

\subsection{Performances at large depths}

\begin{figure}[t!]
    \centering    \includegraphics[width=0.5\textwidth]{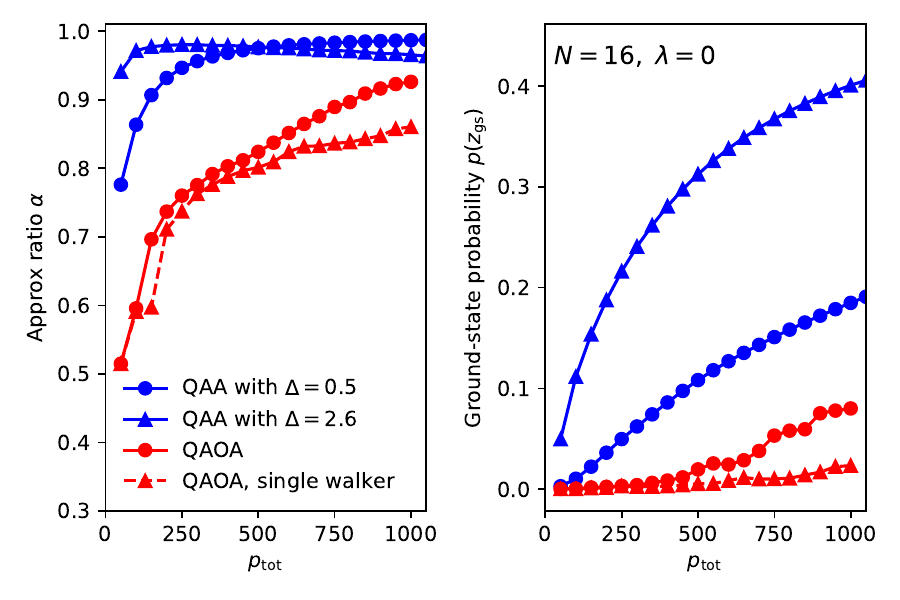}
    \caption{Comparison of the approximation ratio $\alpha$ (left panel) and $p(z_{\rm gs})$ (right panel) between QAOA and QAA. The calculations were performed for 16 nodes using the full statevector in the unconstrained case $\lambda=0$, as a function of the total number of applied layers $p_{\rm tot}$. We show QAA for different choices of the $\Delta$ time parameter, QAOA with multiple walkers, and QAOA with a single walker.}
    \label{fig:approx_ratio_n16}
\end{figure}

In this section, we compare the performances of the QAOA and the QAA as a function of the total number of applied layers $p_{\rm tot}$ defined above, corresponding to the time complexity of each algorithm, for a given subgraph of the antenna problem outlined above for different numbers of sites $N$. the depth $p_{\rm tot}=N_{\rm iter} \cdot p$ with $p$ the single-circuit depth, and $N_{\rm iter}=1$ for QAA and $N_{\rm iter}=50$ for QAOA. In the count of the QAOA number of layers, we do not include the number of walkers used to explore the high-dimensional space of parameters, as they can be run concurrently.

In Figure \ref{fig:approx_ratio_n16}, we show the $\alpha$ obtained using the full statevector with the different methods. The results correspond to the unconstrained case $\lambda=0$. First of all, we notice that QAA with parameter $\Delta=0.5$ (blue circles) steadily but slowly converges to $\alpha=1$, while choosing a value of $\Delta=2.6$ (blue triangles), which is close to the breaking of the Trotter approximation for this problem, leads to a non-monotonic behavior with an initial increase in accuracy followed by a decrease. The QAOA always displays a monotonic behavior as the number of layers is increased, both with a single walker (red triangles) and using several walkers (red circles) to explore the parameter space. As expected, the use of a multi-walker method for augmented optimization allows to find better minima compared to a purely local optimization based on heuristics, especially at larger depths. To summarise, the QAA gives a consistently larger approximation ratio than QAOA at a fixed total number of layers for this problem instance.
If we now turn to the probability of measuring the ground state, we find that similar considerations hold also for that metric, as shown in Fig.\ref{fig:approx_ratio_n16} in the right panel. In particular, we notice that the probability $p(z_{\rm gs})$ for QAA with $\Delta=2.6$ is of approximately $0.4$, which ensures that few measurements are sufficient to find the exact solution with high confidence, and still increases at depths where the approximation ratio starts decreasing.

\begin{figure}[t!]
    \centering    \includegraphics[width=0.5\textwidth]{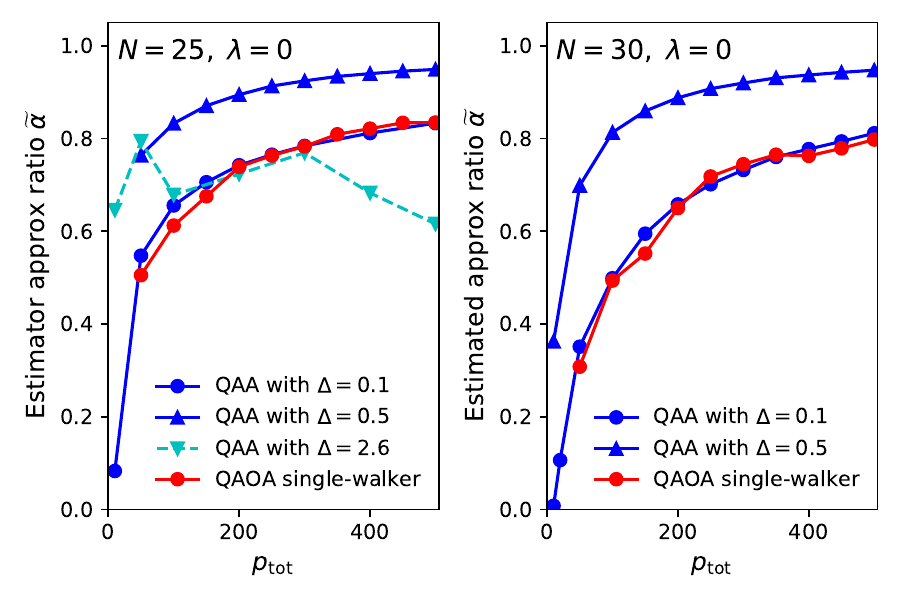}
    \caption{Comparison between QAOA and QAA of the estimator or the approximation ratio $\widetilde{\alpha}$ as a function of the total number of applied layers for a larger number of sites, in the unconstrained case. The left panel shows the case $N=25$, and the right panel shows $N=30$. The number of shots for each calculation is $N_{\rm meas} = 10^4$. The other parameters of the calculations are the same as in Fig.\ref{fig:approx_ratio_n16}.}
    \label{fig:approx_ration_n25}
\end{figure}

In Fig.\ref{fig:approx_ration_n25}, we show the approximation ratio for $N=25$ sites. This problem size is beyond the limit of what we can repeatedly simulate using the full statevector, so we needed to resort to the approach based on measurements with $N_{\rm meas} = 10^4$. Despite the use of a single walker for QAOA, we observe a nice convergence of the estimator $\widetilde{\alpha}$ as $p_{\rm tot}$ is increased. The use of multiple walkers introduces only a marginal improvement on $\widetilde{\alpha}$ for $\lambda=0$ in the range of depths considered here. Depending on the choice of the time parameter $\Delta$, we can obtain a QAA solution that lies close to the QAOA result ($\Delta=0.1$), exhibits a much better convergence to $H_{\rm min}$ compared to QAOA ($\Delta=0.5$), or does not converge at all as we increase the number of time-steps ($\Delta=2.6$). Similar considerations for $\widetilde{\alpha}$ hold when we increase the number of sites to $N=30$. In this case, the execution of QAOA with multiple walkers is prohibitively expensive, so we used only a single walker. Accidentally, the $\alpha$ of QAOA with a single walker overlaps with the results obtained with QAA at $\Delta=0.1$. 

\begin{figure}[t!]
    \centering    \includegraphics[width=0.5\textwidth]{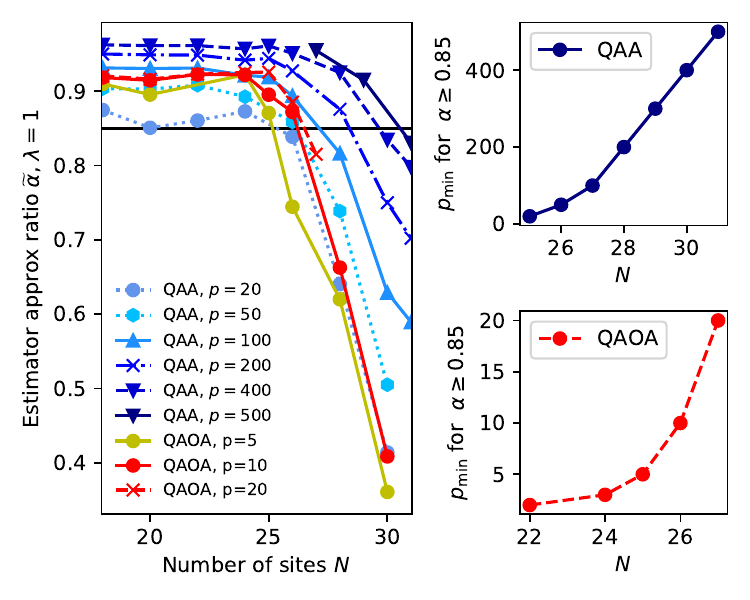}
   \caption{Left panel: Estimator $\widetilde{\alpha}$ as a function of the number of sites $N$ for the constrained case with $\lambda=1$. Curves with different shades of blue indicate QAA results at different depths $p$, while the reddish colors indicate the QAOA results. The black line marks the value $\alpha=0.85$. Upper right panel: minimum value of $p$, named $p_{\rm min}$, that permits to reach $\widetilde{\alpha} \geq 0.85$ in QAA. Lower right panel: same results for QAOA. All the calculations were performed with $N_{\rm meas} = 10^4$ shots.}
\label{fig:p_vs_n}
\end{figure}

In Fig. \ref{fig:p_vs_n}, we analyze the behavior of the estimator of the approximation ratio $\widetilde{\alpha}$ in the constrained problem at $\lambda=1$.
We observe that, at fixed $p$, the estimator $\widetilde{\alpha}$ remains substantially constant up to some threshold $N$ for both the QAA and QAOA algorithms. Above that upper value, the performances quickly degrade, resembling a phase transition in a physical system. It is clearly possible to increase the value of $N$ that can be reached with a given accuracy by increasing the value of $p$. We can also reverse this statement by finding the minimum depth $p_{\rm min}$ that allows to reach a given approximation ratio at each $N$. The answer to this question in the case of QAA is depicted in the upper right panel of Fig. \ref{fig:p_vs_n} for a threshold value $\widetilde{\alpha}=0.85$. The minimum depth $p_{\rm min}$ exhibits a linear behavior in the regime with $27 \leq N \leq 31$, even for the fully connected problem.
The fitted observed linear scaling is $p_{\rm min} \sim 100 N - 2600$, with $N \geq 27$. This means that the depth required to reach the same $\widetilde{\alpha}$ for $N=100$ would be $p_{\rm min} = 7400$, assuming the linear scaling as the asymptotic behavior. We note that at small $N$, $p_{\rm min}$ goes to a constant value $p_{\rm min} \approx 20$ at $\alpha=0.85$ due to the fact that QAA is an adiabatic algorithm and requires a minimum evolution time, hence a minimum number of layers, even for small qubit size $N$.  

For what concerns QAOA, we can see that the $\widetilde{\alpha}$ exhibits a qualitatively similar behavior to QAA. In the lower right panel of Fig. \ref{fig:p_vs_n}, we performed the analysis of $p_{\rm min}$ for QAOA. As expected, the single-circuit depth $p$ to reach a given $\alpha$ is much lower than QAA. The scaling behavior of $p_{\rm min}$ \emph{vs} $N$ is exponential in the range of $N$ considered here. A value of $\alpha \geq 0.85$ could be reached at $p \leq 20$ only up to 27 qubits. The observed exponential scaling of $p_{\rm min}$ in QAOA can either be attributed to intrinsic limitations of the algorithm or to the quality of the local minimum identified by the INTERP-based optimizer in the $(\boldsymbol{\beta}, \boldsymbol{\gamma})$ parameter space. Further studies varying the classical optimizer are needed to investigate possible improvements of the asymptotic behavior.

We point out that, contrary to the $\lambda=1$ case, the step behavior of $\alpha$ as a function of $N$ at fixed depth does not occur in the $\lambda=0$ case, at least in the range of problem sizes considered here. In that case, $\alpha$ seems to decrease continuously and slowly with the system size, suggesting that the full connectivity plays an important role in this performance degradation.

\subsection{Considerations on the probability}

\begin{figure}[t!]
    \centering    \includegraphics[width=0.5\textwidth]{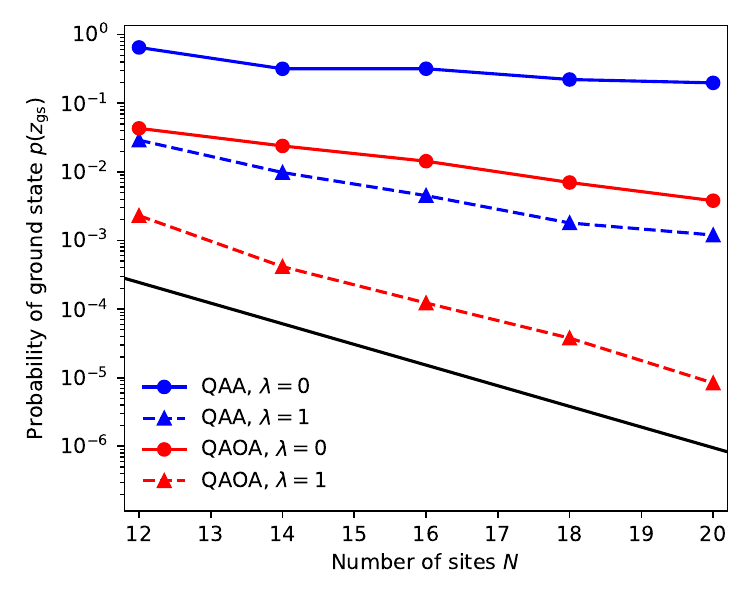}
    \caption{Probability of the ground state for different numbers of nodes. QAOA results are at $p=10$, QAA results at $p=500$. The ground state is always the most probable state in these calculations. The black solid line displays an arbitrary basis state probability of an equal superposition of all the basis states.}
    \label{fig:prob_vs_n}
\end{figure}

Although the approximation ratio $\alpha$ is widely used as an estimation of the accuracy of these algorithms, it only provides an indication relative to the cost function. 
However, it does not provide information on the optimal string $z_{\rm gs}$. This information has to be extracted by a probabilistic sampling of the $|\boldsymbol{\beta},\boldsymbol{\gamma}\rangle$ state in the computational basis.
In Fig.\ref{fig:prob_vs_n}, we show the value of this probability $p(z_{\rm gs})$ varying $N$ for the constrained and unconstrained cases with QAA and QAOA. These results have been obtained using the full statevector emulator. The depth of the QAOA circuit was chosen to be $p=10$. In the QAA case, we choose a depth $p=500$, so that the total amount of applied gates is equal to QAOA, $\Delta=2.6$ for $\lambda=0$ and $\Delta=0.4$ for $\lambda=1$, as these choices result in high values of $\alpha$. The results are compared to the case of a flat state in which all states are equally possible (black solid line). 

As we can see, in the unconstrained $\lambda=0$ case, the probability of measuring the ground state is larger in QAA compared to QAOA of an order of magnitude. However, in both algorithms, $p(z_{\rm gs})$, decreases noticeably as the number of qubits decreases. Crucially, the decrease of $p(z_{\rm gs})$ $vs$ the system size $N$ is exponential within the considered range. At fixed $p_{\rm tot}$, the QAA exhibits a smaller slope than QAOA. In the constrained case with $\lambda=1$, the $p(z_{\rm gs})$ is a couple of orders of magnitude smaller for both algorithms. The numerical coefficients of the exponential scaling are summarized in Table \ref{tab:my_label}. Considering that we need $N_{\rm meas} \sim 1/p(z_{\rm gs})$ measurements on average to sample the ground state, it becomes quickly impractical to measure it at least once as the system size increases.

\begin{table}[]
    \centering
    \begin{tabular}{|c|c|c|}
    \hline
    Considered case & Algorithm & Average $N_{\rm meas}$ \\
    \hline
      $\lambda = 0$ & QAA  & $1.07^N$ \\
      $\lambda = 0$ & QAOA  & $1.30^N$ \\
      $\lambda = 1$ & QAA & $1.36^N$ \\
      $\lambda = 1$ & QAOA & $1.67^N$ \\
      \hline
    \end{tabular}
    \caption{Scaling of the inverse probability of measuring the ground state as a function of $N$ for the different cases considered here. This corresponds to the minimum $N_{\rm meas}$ needed to measure the exact solution at least once on average. The parameters are $p=10$ for QAOA and $p=500$ for QAA.}
    \label{tab:my_label}
\end{table}

\subsection{NISQ case: a limited amount of measurements and low depth}

In real devices, the number of applicable gates is currently limited to a few thousand and only in cases when the problem can be directly mapped on the quantum processor \cite{Kim2023}. For instance, in a recent IBM experiment that aimed to show the utility of NISQ quantum computers, the number of applied gates was $|G|=2,880$ \cite{Kim2023}. The authors managed to extract expectation values using error mitigation techniques with high accuracy \cite{Kim2023}. Considering the average connectivity $\langle c_\lambda \rangle$ of our problem, we know that $\langle c_0\rangle \approx 5$ and $\langle c_{\lambda > 0} \rangle = N-1$, that correspond to a number of two-qubit gates per layer $|G^{(2)}|_{\lambda = 0} \approx 5 N$ and $|G^{(2)}|_{\lambda > 0} \approx  N (N-1)/2$. The number of single-qubit gates per layer is $|G^{(1)}| = 2N$. Assuming optimistically that the current maximum number of gates sequentially applicable on a NISQ device with the number $|G|$ proposed by IBM in their experiments, and assuming that the connectivity of the problem fits on the hardware without the need for swap gates, we obtain that the maximum number of antennas $N_{\rm max}$ that can be simulated at depth $p=1$ on this hypothetical device is $N_{\rm max} = 425$ for $\lambda = 0$ and $N_{\rm max} = 75$ for $\lambda > 0$. Conversely, if we fix $N=100$ potential sites, we can reach depth $p_{\rm max} = 4$ for the unconstrained case, while the constrained case does not fit in the machine. We need $|G| \approx 6000$ to execute the constrained case with $N=100$ with depth $p = 1$ on this hypothetical hardware.

\begin{figure}[t!]
    \centering    \includegraphics[width=0.5\textwidth]{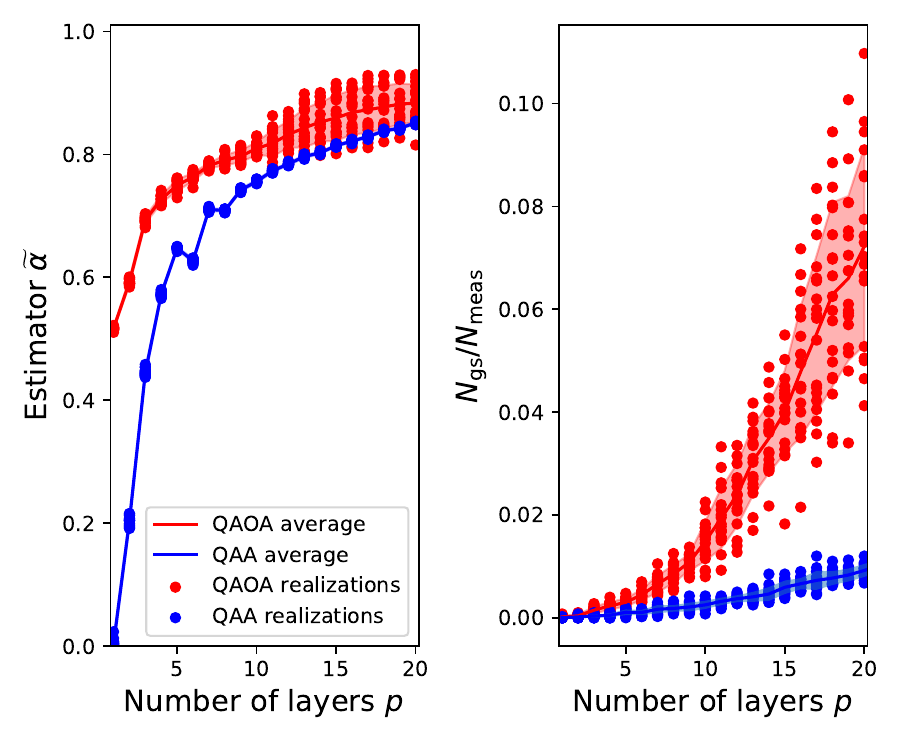}
    \caption{Comparison of the estimator of the approximation ratio $\widetilde{\alpha}$ (left) and the normalized counts for QAA (blue) and QAOA (red) as a function of the number of layers $p$, with $N_{\rm meas} = 4000$ and $N=$16, $\lambda=0$. The circles indicate single realizations obtained by varying the random seed for the measurement set, while the solid lines and the shaded area indicate respectively the average and standard deviation obtained aggregating 15 different random seeds. }
    \label{fig:finite_meas}
\end{figure}

Due to these considerations, it is also crucial to compare the performances at the fixed depth $p$, i.e. the number of consecutive layers applied in a single circuit execution. In this way, we can assess the algorithmic performance assuming the expected limitations associated to an implementation on a device of the current generation. 
In Figure \ref{fig:finite_meas}, we compare the outcome of QAOA and QAA at a finite number of measurements $N_{\rm meas}=4000$ and using a depth up to $p=20$. The case shown here has $N=16$ antennas and $\lambda=0$ (unconstrained case). Each calculation is repeated 20 times with different random seeds for the measurements,  generating the sequence of INTERP starting points independently for each seed.  
We first consider the $\widetilde{\alpha}$ estimator, shown in the upper panel. QAOA reaches much higher performances than QAA when $p < 10$ while, above that threshold, both the QAOA and the QAA reach comparable values. Nonetheless, QAA does so without optimization of the parameters $\boldsymbol{\beta}$ and $\boldsymbol{\gamma}$. 
In the lower panel, we show the ground-state counts $N_{\rm gs}/N_{\rm meas}$. We see that QAOA exhibits better performances on average, but also a much higher variance between different repetitions compared to QAA. The high variance is related to the optimization procedure in higher dimensions, which finds different local minima for each replica. Finding the minimum in the landscape contributes to the variability of the result more than shot noise arising from limited measurements in this case. 

\begin{figure}[t!]
    \centering    \includegraphics[width=0.5\textwidth]{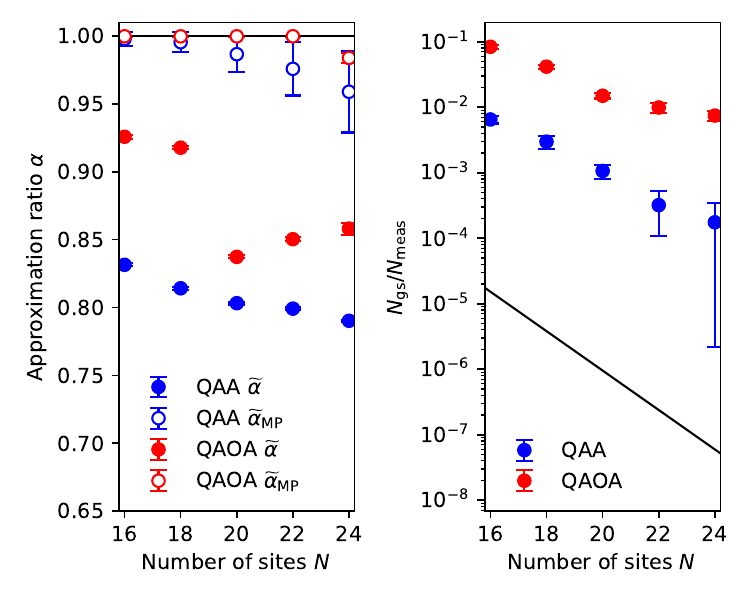}
    \caption{Performances of the QAA (blue) and QAOA (red) algorithms with $N$ sites/qubits at $p=20$ in the unconstrained case with $\lambda=0$. Circles and error bars respectively indicate the average and the standard deviation over 20 samples with $N_{\rm meas}=10^4$ shots each. Left: $\widetilde{\alpha}$ and measured most-frequent state $\widetilde{\alpha}_{\rm MP}$ (empty circles). The black line represents the exact $\alpha=1$. Right: Normalized number of counts. Here the black line is $2^{-N}$ (probability of a string in an equal superposition of states). }
    \label{fig:finite_meas_vs_n}
\end{figure}

\begin{figure}[t!]
    \centering    \includegraphics[width=0.5\textwidth]{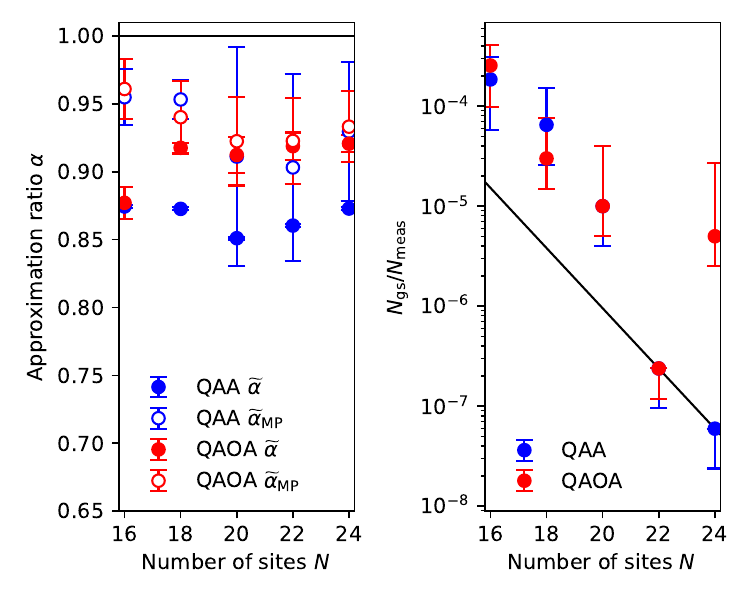}
    \caption{Performances of the QAA (blue) and QAOA (red) algorithms with $N$ sites/qubits at $p=20$ in the unconstrained case with $\lambda=1$. The details of the calculations and the explanation of the legend is the same as in Fig. \ref{fig:finite_meas_vs_n}.
    }
\label{fig:finite_meas_vs_n_const}
\end{figure}

At this point, we can also investigate the fixed-shot approximation ratio and counts, as the system size increases. In Fig.\ref{fig:finite_meas_vs_n}, we show the $\lambda = 0$ case, while in Fig.\ref{fig:finite_meas_vs_n_const}, we show the $\lambda = 1$ case. Each point is the result of 20 calculations with different random seeds for the measurements, and all the QAOA optimizations were initialized with the same parameters, previously obtained with INTERP, hence the variance is comparable to QAA. The depth of both QAA and QAOA is fixed at $p=20$. The number of measurements is kept fixed to $N_{\rm meas}=10^4$. In both figures, the left panels show the value of $\tilde{\alpha}$ (full circles) and $\tilde{\alpha}_{\rm MP}$ (full circles) for QAA (blue markers) and QAOA (red markers). Instead, the right panels show the normalized counts $N_{\rm gs}/N_{\rm meas}$ for the two algorithms. For $\lambda=0$, we observe that $\alpha$ is higher for QAOA compared to QAA. QAOA also manages to retrieve the ground state as the most-frequent state, as signalled by $\tilde{\alpha}_{\rm MP} = 1$, while at this depth QAA shows a decrease in $\widetilde{\alpha}_{\rm MP}$ as $N$ increases. The $N_{\rm gs}/N_{\rm meas}$ metric is one order of magnitude higher in QAOA. However, even though the ground state has a relevant weight in both QAOA and QAA, they both show an exponential decrease of $N_{\rm gs}/N_{\rm meas}$ as the system size increases. 

If we consider the $\lambda = 1$ case, we realize that the constrained problem is much harder even for the quantum algorithm. First of all, the $\widetilde{\alpha}$ is comparatively larger than for $\lambda =0$, but $\alpha_{\rm MP}$ is lower which means that the most-frequent state almost never coincides with the ground state at $N \geq 16$. Additionally, the probability $p(z_{\rm gs})$ is much smaller in the constrained case, as shown in the previous section. This fact is reflected in the $N_{\rm gs}/N_{\rm meas}$. Additionally, the variance is very high and, for larger $N$, the ground state is not measured in most of the realizations.
We observe that the ground state usually appears in the sampling up to $N=20$ at least once for both choices of $\lambda$. Above that threshold, we observe an increasingly small probability of measuring $z_{\rm gs}$ in a given sample with $N_{\rm meas} = 10^4$.
This is a computational demonstration of the fact that an observable such as $\alpha$ is much less susceptible to shot noise than the probability of each single string, since it is averaged over all the sampled values.

\subsection{Quantum algorithms as generators of suboptimal strings}

\begin{figure}
    \centering
    \includegraphics[width=0.5\textwidth]{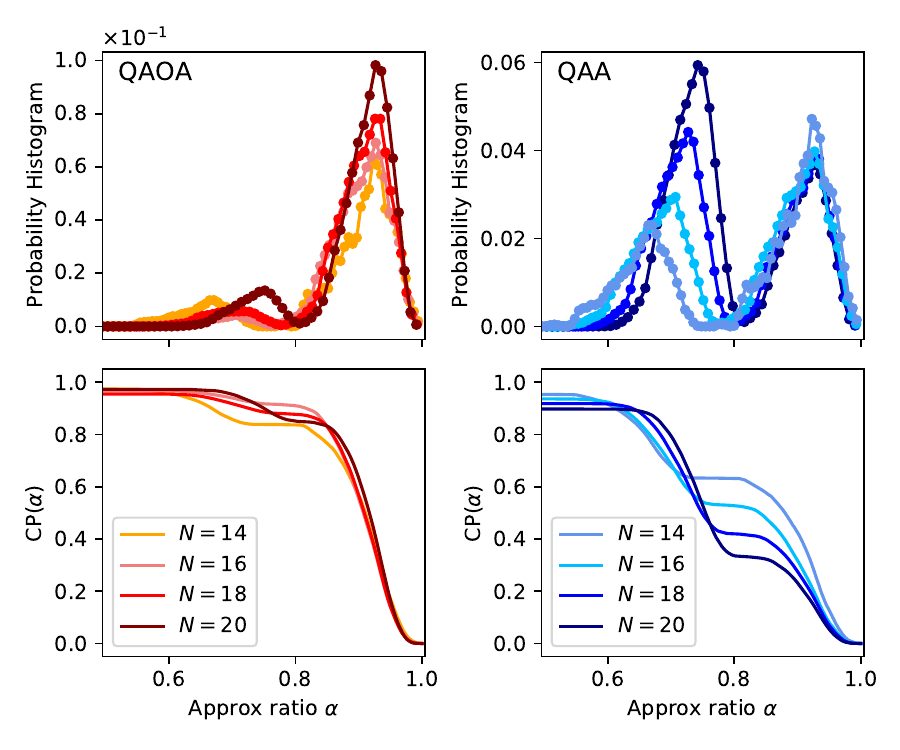}
    \caption{Probability histogram (top row) and cumulative distribution (bottom row) of the generated strings ordered as a function of their approximation ratio $\alpha$. The bins are $0.01$ wide. Both the QAOA result (left panels, reddish colors) and QAA (right panels, bluish colors) are obtained at low depth $p=5$ and $\lambda=1$. Different colors indicate a different number of sites $N$. The probabilities are extracted from the exact statevector.}
    \label{fig:cumul_prob}
\end{figure}

The two quantum algorithms discussed here return a probability distribution which is peaked at low cost at sufficiently large $p$. A more peaked distribution allows for a smaller number of measurements. The availability of only a limited number of measurements $N_{\rm meas}$ introduces an additional difficulty: $z_{\rm gs}$ could not appear in our sample simply because of statistical fluctuations and even if its probability $p(z_{\rm gs})$ is the highest among all $p(z)$s. In our simulations, this is often the case if $N_{\rm meas} \ll 2^N$, as shown in the previous paragraph (Figs.\ref{fig:finite_meas_vs_n}-\ref{fig:finite_meas_vs_n_const}). 

In the majority of industrial applications, it is not necessary to find the \emph{exact} result $z_{\rm gs}$, but it is sufficient to find an acceptable configuration $z_{\rm good}$, close enough to $z_{\rm gs}$ according to some criterion. In the case of the antenna network model considered here, there can be some tolerance in avoiding the overlaps or in the allocation of the local resources. If we introduce a minimum allowed approximation ratio $\alpha$, any configuration $z_k$ with approximation ratio $\alpha_k \geq \alpha$ is an acceptable configuration. This introduces an acceptable cost window $\delta H = H_{\rm min} (1 -\alpha)$. We can consider the probability to measure any arbitrary string satisfying this condition by computing the cumulative probability ${\rm CP}(\alpha) = \sum_{\alpha_k \geq \alpha} p(z_k)$. Correspondingly, in the case of limited shots, we can compute its estimator $\widetilde{{\rm CP}}(\alpha)$ obtained by replacing $p(z_k)$ with the normalised counts.

In Fig.\ref{fig:cumul_prob}, we show the exact probability histogram and ${\rm CP}$ for QAOA and QAA at different problem sizes. We plot the histogram with bins of width $0.01$, since for these problem sizes plotting all the $2^N$ probabilities results in a very packed and uninformative picture. This corresponds to averaging $p(z)$ within the bin width. As expected, the probability histogram is peaked close to the ground state ($\alpha=1$) and ${\rm CP}$ decreases as we approach values of $\alpha$ close to unity.  
In the constrained case $\lambda=1$, the most-probable state $z_{\rm MP}$ coincides with $z_{\rm gs}$ already at depth $p=5$, even for QAA. Nevertheless, the probability of the ground state $p(z_{\rm gs})$ itself is very small already for $N=20$: $p(z_{\rm gs}) \approx 10^{-4}$ for QAOA and $p(z_{\rm gs}) \approx 2.4 \cdot 10^{-6}$ for QAA. However, if we allow a minimum $\alpha$, the probability of measuring a string with an approximation ratio at least $\alpha$ grows rapidly. This strong increase appears for all the choices of $N$ in the range in which we can access the full probability distribution. Problems with large $N$ show a sharper increase as $\alpha$ is reduced, as reflected in the higher maximum of the probability histogram. This is due to a higher density of strings. Our results suggest that we can measure an acceptable string $z_{\rm good}$ with high probability using a polynomial number of measurements $N_{\rm meas}$ as the problem size is increased. 

\begin{figure}
    \centering
    \includegraphics[width=0.5\textwidth]{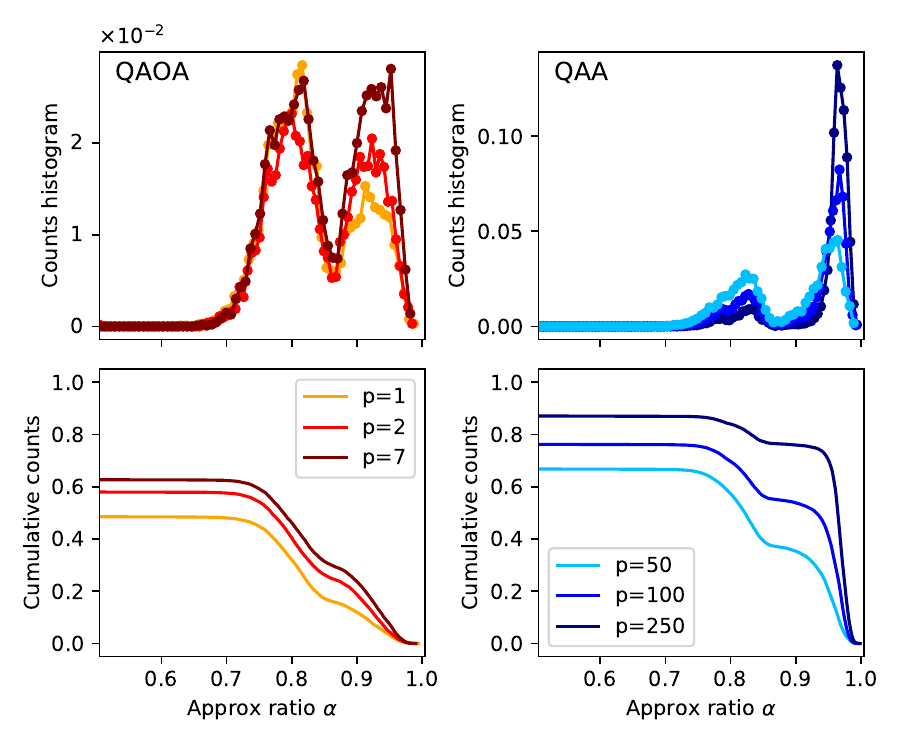}
 \caption{Counts of the measured strings with $N_{\rm meas}=10^4$ for $N=30$. The horizontal axis represents the approximation ratio $\alpha$. The results are obtained with QAOA (left in reddish colors) and QAA (right, blue colors) in the constrained case with $\lambda=1$. Different colors indicate a different depth $p$. Top panels: histogram of the counts, using dots for visibility (lines are guides for the eye). The bins are 0.01 wide. Bottom panels: cumulative counts $\widetilde{\rm CP}$.}
    \label{fig:cumul_prob_n30_qaa}
\end{figure}

In Fig.\ref{fig:cumul_prob_n30_qaa}, we show that this procedure can be applied to a moderately large network with $N=30$ for both QAOA and QAA. Even though the ground state is never measured in case with $N_{\rm meas} = 10^{4}$, we can still identify strings with $\alpha$ close to 1 with rather high probability as signalled by the $\widetilde{\rm CP}$ estimator. We notice that QAOA displays two separate peaks with different numbers of operating antennas. QAA at large depths exhibits an extremely sharp peak centered at costs very close to the ground state, while at moderate $p$ it exhibits a much broader distribution with two equally high peaks resembling that of QAOA. The vicinity of a peak to the ground state and its width are controlled by the depth $p$, no matter if we use QAA or QAOA. In fact, our results show that the optimization of the QAOA parameters helps in reducing the minimum $p$ required to get a well-defined peak. As a consequence, the probability of measuring low-cost strings close to $\alpha=1$ increases with $p$.
As in the case of classical state-of-the-art approximate solver, the utility of this approach strongly depends on the minimum $\alpha$ allowed for the specific use-case of interest. Some practical applications may need extremely narrow windows, such that the corresponding cumulative probability is still too low to be reached with practical $N_{\rm meas}$ and $p$ for large problem sizes. In general, however, quantum algorithms appear to effectively suppress the weight of high-cost strings, which enhances the probability of measuring $z_{\rm good}$ strings characterized by low costs.

\section{Conclusion}

We conducted a comparative analysis of the QAOA and QAA algorithms for the antenna placement problem. 
We implemented two model variants, one with constraints ($\lambda \neq 0$) which imply full connectivity of the underlying graph, and the other without constraints ($\lambda=0$) which features local connectivity only. 
The variants we considered can be considered representative of a wide range of potentially relevant industrial applications. Hence, our work provides a useful comparison and reference for other combinatorial problems on general graphs.

The work is carried out with exact statevector emulation using high-performance computing resources. These enabled the numerical study of system sizes up to 31 qubits, covering a range not frequently discussed in the literature. 

We evaluated the algorithm performances, using three different metrics: the approximation ratio $\alpha$, the ground state probability $p(z_{\rm gs})$, and the cumulative probability ${\rm CP}$. 
The latter two metrics are considered to evaluate the quality of the specific candidate solution strings that are needed to take effective actions in real-world scenarios, while $\alpha$ provides an estimate of how close the expectation value of the cost function is to the exact solution. 

We started our analysis by comparing the $\alpha$ of QAA and QAOA for our specific problem instances at a fixed total number of applied layers $p_{\rm tot}$. This corresponds ideally to fixing the execution time. In this setting, QAA consistently exhibits better performances than QAOA. Conversely, if we fix the single-circuit depth $p$ and work in the low-depth NISQ regime, QAOA outperforms QAA. 

These results confirm that QAA is preferable to QAOA when fault-tolerant quantum computers become available, while QAOA is more suitable for execution on NISQ devices. The successful implementation of the QAOA algorithm remains however limited by the need for high-dimensional optimization in the parameter space.

We also numerically estimate how the minimum depth $p_{\rm min}$ varies as a function of $N$ to ensure solutions with $\alpha$ above a predefined threshold. This provides a rough indication of the minimum gate count required to achieve high solution qualities on a hypothetical quantum device with full connectivity and shot error only. More specifically, when a large number of sites $N$ is considered, we observe that $p_{\rm min}$ increases linearly and exponentially in the QAA and QAOA cases respectively. The QAOA scaling might however depend on the chosen optimization strategy.

Our analysis also addressed quality of the generated strings.  
Here, we find that both algorithms predict the most-probable state $z_{\rm MP}$ to coincide with $z_{\rm gs}$ at sufficiently large depths. At the same time, the probability $p(z_{\rm gs})$ is found to decrease exponentially with the problem size, which in turn implies an exponential $N_{\rm meas}$ to extract $z_{\rm gs}$ in large instances. This makes the identification of $z_{\rm gs}$ impractical with a limited $N_{\rm meas}$, especially in the constrained case.
Besides, the calculated ${\rm CP}$ indicates that QAOA and QAA can be used to generate approximate solutions close to the exact one with high probability. This is suggested by the fact that the ${\rm CP}$ of solutions in a given reasonable window of cost does not significantly depend on the size in the considered range. 

From a practical standpoint, the required circuit depths and the large number of shot measurements, associated with the low ground-state probability, are recognized as limiting factors for the successful application of these algorithms to solve real-world use-cases \emph{exactly}. 
Several strategies to mitigate these issues are a matter of on-going research, and will be addressed in future analyses of the problems considered here. 
Nevertheless, our analysis clearly indicates that QAA and QAOA can still be successfully applied to generate high-quality approximate solutions, in a similar way as quantum annealers and widely-used heuristic classical solvers. In these regards, our findings provide evidence that these and other gate-based quantum algorithms have the potential to be profitably deployed at scales relevant to industrial use-cases.

\FloatBarrier

\bibliography{main}

\end{document}